\documentclass{aa}
\usepackage{epsfig}

\newcommand{\msun}{M$_{\odot}$}

\newcommand{\degree}{$^{\rm o}$}

\newcommand{\bb}{\bibitem[]{bla}}

\def\lesssim{\mathrel{\hbox{\rlap{\hbox{\lower4pt\hbox{$\sim$}}}\hbox{$<$}}}}
\def\gtrsim{\mathrel{\hbox{\rlap{\hbox{\lower4pt\hbox{$\sim$}}}\hbox{$>$}}}}

\def\arcsec{\hbox{$^{\prime\prime}$}}

\begin{document}


\title{On the origin of the X-ray emission towards the early Herbig Be star MWC~297}

\author{Jorick S. Vink\inst{1} 
\and Paul M. O'Neill\inst{1} 
\and Sebastian G. Els\inst{2}
\and Janet E. Drew\inst{1}
}

\institute{Imperial College London, Blackett Laboratory, 
           Prince Consort Road, London, SW7 2AZ, U.K.
            \and
           California Institute of Technology, Mail Code 102-8, 
           1200 E.California Blvd., Pasadena CA 91125, USA}

\offprints{Jorick S. Vink, jsv@astro.keele.ac.uk}

\titlerunning{The X-ray emission towards MWC~297}
\authorrunning{Jorick S. Vink et al.}

\abstract{We present high resolution ($\simeq$ 0.2\arcsec) 
AO-corrected coronographic near-infrared imaging on the early-type Herbig Be star MWC~297. 
X-ray flaring has been reported towards this young 
object, however this has been difficult to reconcile with its early spectral type (B1.5) and relatively high 
mass ($\sim$ 10\msun). Our infrared and X-ray analysis shows 
that the X-ray flaring is likely due to a late-type star in the same field. 
The case of MWC~297 emphasizes the need for coronographic imaging to address the reality of X-ray emission 
towards Herbig Ae/Be stars, which is needed to understand the differences 
between low and high-mass star formation.
}

\maketitle


\section{Introduction}
\label{s_intro}

Herbig Ae/Be stars are pivotal young stellar objects for understanding 
the physical differences between low and high mass star formation, as their 
masses ($\simeq$ 2 -- 15 \msun) make them transitional between the low-mass 
T Tauri and massive stars.

X-ray emission is frequently seen in early-type O stars, where 
the X-ray emission arises from shocks in radiatively-driven winds (Lucy \& White 1980), but 
the X-ray emission ceases for B stars. 
On the cool side of the 
Hertzsprung-Russell Diagram (HRD), late-type stars are magnetically active due
to the presence of convective outer layers (at $\sim$ G and later types).
In between these two extremes, B and A stars are not 
considered to be magnetically active (except Am stars), and are 
generally X-ray quiet. 

A different story might hold for the pre-main sequence (PMS) Herbig Ae/Be stars. 
Zinnecker \& Preibisch (1994) and Damiani et al. (1994) found that 30 -- 50 \% of 
the Herbig stars surveyed with the {\sc rosat} and {\sc einstein} satellites 
emit copious amounts of X-rays. Although A stars do not have convective outer layers, 
there may be magnetic activity due to a seed magnetic field. 
In fact, the magnetic accretor model --  
generally applied to lower mass PMS T Tauri stars -- may also be  
at work in Herbig Ae stars (Vink et al. 2002, 
Hubrig et al. 2004). This might be consistent with the X-ray flaring seen 
in some Herbig Ae stars (e.g. Giardino et al. 2004), but whether this scenario
may hold for earlier type objects is questionable given 
the different circumstellar properties of the Herbig Ae and 
Be stars (Vink et al. 2002, Eisner et al. 2003). It is therefore surprising 
that X-ray flaring has been discovered in the massive Herbig Be star 
MWC 297 with the {\sc asca} X-ray satellite (Hamaguchi et al. 2000, 2005).
As MWC~297 has a spectral type as early as B1.5 (Drew et al. 1997), this 
represents a significant puzzle to early stellar evolution theory 
for intermediate-mass ($\sim$ 10 \msun) B stars, and raises the question whether new physical 
effects in this part of the HRD need to be identified.

In surveys of Herbig X-ray emission (Zinnecker \& Preibisch 1994, Damiani et al. 1994), 
X-ray luminosities ($L_{\rm X}$) were found to be correlated with bolometric luminosity ($L_{\rm bol}$): 
as a result the X-rays were attributed to wind shocks -- similar to the O star mechanism. But
Herbig Ae/Be winds are probably not fast enough for radiatively-driven wind shocks to develop.
At the same time, the option of unresolved companions was considered unlikely in view of the
$L_{\rm X}$ versus $L_{\rm bol}$ relationship. However, Testi et al. (1998) have 
found that the number of late-type companions to Herbig Ae/Be stars 
grows with increasing mass and $L_{\rm bol}$. This re-opens the possibility that 
at least some of the X-ray emission towards Herbig stars is due to one or more 
late-type companions. 

In this paper, we address the question of whether the X-ray flaring of MWC~297 
originates from MWC 297 itself or from a late-type companion.
Although many sources around Herbig Ae/Be stars have been 
imaged at separations of 100 -- 10000 AU (Li et al. 1994, 
Pirzkal et al. 1997, Leinert et al. 1997, Smith et al. 2005), companions around the 
more luminous objects are much more challenging to discover. 
Most imaging studies of Herbig stars identify objects at 
relatively large distances ($\simeq$ 10000 -- 100000 AU) from the central object (e.g. Testi et al. 1998).
The fact that no sources have been detected close to MWC~297, does not 
mean that they are not there, especially since MWC 297, with an $H$-band magnitude of 4.4, may 
completely outshine any close-in companions. 

Progress may be made through Adaptive Optics (AO) coronographic 
imaging -- a technique that has only recently become available.
At X-ray wavelengths, early studies with satellites such as {\sc rosat}, {\sc einstein}, and {\sc asca} 
were generally subject to relatively poor spatial resolution, but the 
{\sc chandra} satellite offers spatial resolution of the order of 1$\arcsec$. {\sc chandra} 
might well be capable of resolving many putative low-mass companions, as exemplified by the study of  
Stelzer et al. (2003) on late-type B stars. Here we report 
combined infrared AO-coronographic imaging with high resolution X-ray imaging 
of the extreme case of MWC~297.


\section{Observations}
\label{s_obs}

\subsection{Coronographic Adaptive Optics}

The AO-corrected $H$-band imaging data were obtained during the 
night of 2004 July 31 with the Optimised Stellar Coronograph OSCA (Thompson et al. 2003).
included in the NAOMI AO system (Myers et al. 2003) on the 4.2-metre 
William Herschel Telescope (WHT), La Palma. 
We used an occulting mask of 1.0\arcsec\ to block the light of the 
central object. Using this mask, it is possible to suppress 
the bright core of the stellar point-spread function (PSF) and lower its 
wings by $\Delta H \sim$ 6.5 mag for distances from the 
central star of $\sim$1\arcsec. The NAOMI PSF also includes the 
diffraction pattern of the segmented deformable mirror (DM). 
We took sets of observations under various position angles (PAs of 0, 30, 60 and 120 degrees),  
as the diffraction pattern remains constant over the chip this allows us to also 
inspect areas that would otherwise remain hidden in the diffraction pattern. 
We obtained 2$\times$ 10 sets of 20 co-averaged images with an exposure time 
of 1 sec for each PA. 
In addition, we imaged stars with no known companions as 
PSF references after each PA set.
The atmospheric conditions were good and 
the AO corrected images yield a FWHM of $\sim$ 0"2. 
The centring of the mask over the star was done manually. 
Basic data reduction included dark, and flatfield subtraction as well as 
bad pixel removal, which was performed using the ESO software packages {\tt eclipse} 
and {\sc midas}. 

\subsection{X-ray observations with {\sc chandra} and {\sc asca}}


{\sc chandra} observed MWC~297 in a single visit (observation ID 1883)
on 2001  September 21  and 22, for $\sim$11 hours, using 
the default ACIS-I chips (I0-I3, S2, S3). An aspect 
correction\footnote{http://cxc.harvard.edu/cal/ASPECT/fix\_offset/fix\_offset.cgi}
was  applied to  the level~1  events file,  such that  the 90\% confidence 
error circle has  a radius of  0.6\arcsec.  New  level~2 events
files were created  in the standard manner using  CIAO 3.2.1 and CALDB
3.0.1. The resulting good exposure time was $\sim$37~ks.
A 0.3--10~keV  image was extracted and source  detection was performed
using  the CIAO  tool  {\tt wavdetect}. Source  counts were  extracted
using a circular region centred on each source, with a radius of either
3  or 5  pixels  (see Section~\ref{s_res2}).   Background counts  were
extracted from an annulus with a width of 100
pixels centred on the source aperture. The inner radius of the annulus was set to be twice the source
region radius.   An exposure map was  generated, and this  was used to
determine  the  scaling  factor  between  the  source  and  background
regions. The estimated mean number of background counts in each of the
source regions is $\sim$0.4~counts.
%

{\sc  asca} observed MWC~297 three times (with sequence numbers 21007000,
21007010, 21007020) between 1994 April 8 and 12.  These data were presented 
previously by Hamaguchi et~al. (2000,2005). Here we only discuss the 
$\sim$6~hour observing sequence 21007010, during which the decay phase of an X-ray flare 
was observed.
Moreover, we restricted our analysis to the data from the 
two Solid-state Imaging Spectrometers (SIS), SIS0 and SIS1, as these 
instruments have higher spatial resolution than the Gas 
Imaging Spectrometers on {\sc asca}.
The PSF of the SIS instruments have a core with a FWHM of $\sim$50\arcsec\
(Jalota et al. 1993).
The data were screened using the Tartarus\footnote{http://astro.imperial.ac.uk/Research/Tartarus/} 
analysis pipeline, which yielded a good exposure 
time of $\sim$4.7~ks for each SIS instrument.
Sky images in the 0.5--10~keV band were extracted for each SIS using a
spatial binning of 1.6\arcsec\ per pixel. These images were  then combined
and smoothed with    a     Gaussian     kernel    having     a
FWHM of 4 pixels.
The source position calibration of the `REV2' data in the {\sc asca}
archive  has  a  90\% confidence error 
circle with radius $\sim$ 40\arcsec\
(Gotthelf 1996).
A later calibration allows this pointing accuracy to  be improved, with
the 90\% confidence error being reduced to a radius of 12\arcsec. We
used the on-line look-up
table\footnote{http://heasarc.gsfc.nasa.gov/docs/asca/coord/update.html}
to restore the astrometric accuracy
of our combined SIS0+SIS1 image.


\section{Results}
\label{s_res}

\subsection{Coronographic Adaptive Optics}
\label{s_res1}

\begin{figure}
\centerline{\psfig{file=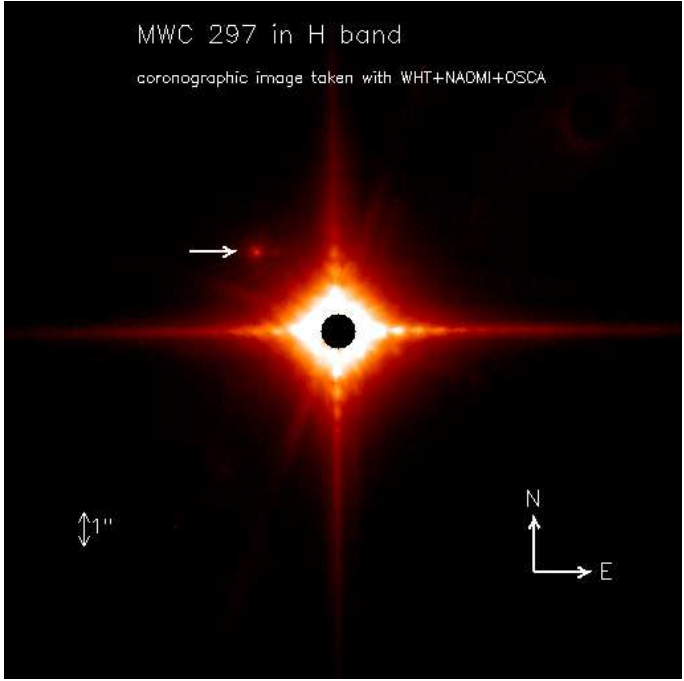, width = 9 cm, angle=360}}
\caption{
OSCA $H$-band image of MWC~297. The logarithmically-scaled image shows the
presence of a faint source close to the bright star, at PA=313\degree.
In order to increase the visibility of this object we placed
a numerical mask over the central region which is covered by the
instrumental coronographic mask.}
\label{f_osca}
\end{figure}

Figure~\ref{f_osca} shows the OSCA $H$-band image of MWC~297. 
We report the discovery of a faint object located 3.39\arcsec\ $\pm$ 0.2\arcsec\ along position 
angle (PA) 313\degree\ $\pm$ 2\degree\ (northwest) from MWC~297. Given the fact 
that the newly found source rotates with the 
rotator position, we consider the detection to be real. 
Some other features are visible, in particular a large 
cross-like pattern extending over almost the entire field and 
two fainter spikes at PAs of $\sim$ 70\degree\ and 250\degree. The cross-like 
feature is the DM's diffraction pattern, whereas the latter is 
due to the secondary spider. 
Finally, a small fainter structure can be seen just east of the newly detected
source. As its position does not change on the sky in accordance with the rotator
angle, this structure is considered to be 
an instrumental artifact.
 
Although we have not proven a physical relationship
between  MWC~297 and the newly discovered object, the projected separation 
of 850 AU (distance to MWC~297 $=$ 250 pc; Drew et al. 1997) is consistent with the new 
source being a binary companion or to have formed from the same cloud core as MWC~297 (Li et al. 1994).
 
Accurate photometry of the newly found object is challenging as 
it is found close to the bright host star. 
Using the 2MASS $H$-band magnitude of MWC 297, we calibrated our zeropoint 
by using open-loop images of MWC 297. 
PSF photometry on the companion was performed by fitting Gaussians. 
In addition, we performed aperture photometry. 
The resulting $H$-band magnitudes agree fairly well with each other, and we find
an $H$-band magnitude difference of $\Delta H$ = 8.5 $\pm$ 0.25 mag. 
Using the 2MASS magnitude of MWC~297, we find $H$ = 12.9 $\pm$ 0.25 mag for the {\sc osca} source.
To investigate whether our discovered object at a PA of 313\degree\
could be responsible for the X-ray flaring of MWC~297, we turn to the
relevant archival X-ray satellite data.

\subsection{{\sc chandra} and {\sc asca} results}
\label{s_res2}

\begin{figure*}
\centerline{\psfig{file=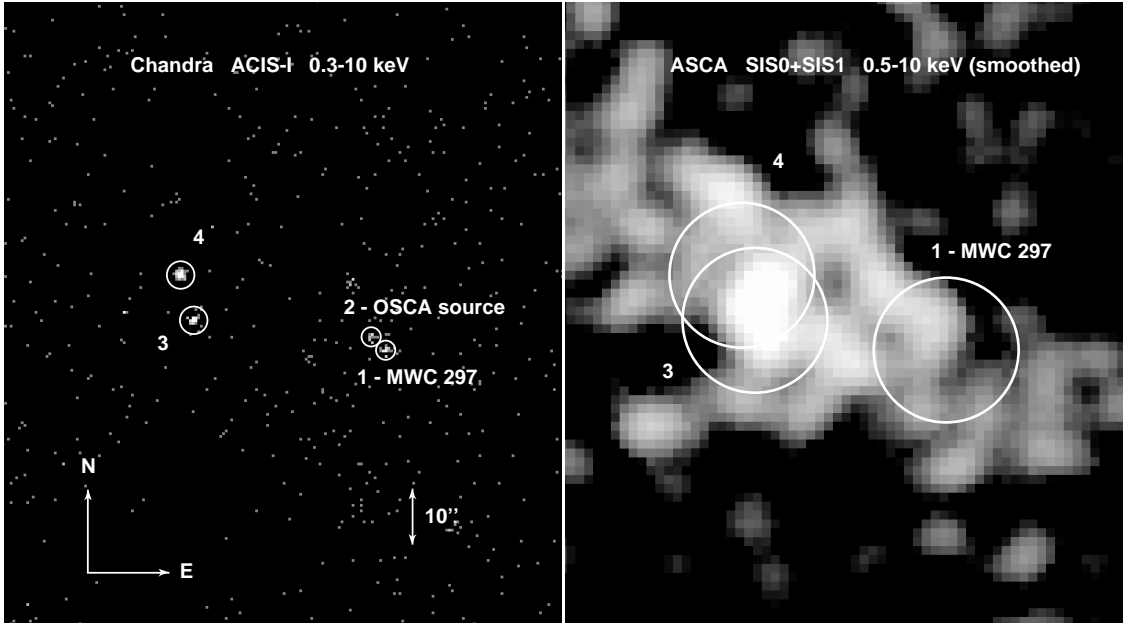, width = 15 cm}}
\caption{{\sc  chandra} (left) and {\sc asca} (right)  images of  the
MWC~297  field.   Note  the  presence  of two  X-ray  sources  in  the
{\sc chandra}  image, with  a separation  of  $\simeq$ 3.5\arcsec\ at PA  
of $\simeq$ 315\degree. The {\sc asca} image (same scale) is from  the observation
during which the source was seen  to flare, and clearly shows the much
larger  PSF  of {\sc asca}.   The  positions  of  the {\sc chandra}
sources 1 (MWC~297), 3, and 4  are shown, along with the error circles.
Note that, in  the {\sc asca} image, the observed peak in intensity is
consistent with positions of sources 3 and 4, but not with that of MWC~297.}
\label{f_xray}
\end{figure*}

The {\sc chandra} X-ray image of the region near MWC~297 is shown  in
the left panel of Fig.~\ref{f_xray}, where we note four objects. 
The positions of these objects are listed in 
Table~\ref{t_chan}. The two X-ray sources with a separation 
of $\simeq$ 3.5\arcsec\ at PA of $\simeq$ 315\degree\ are an
exact match to our $H$-band imaging  of MWC~297 (source 1) and our
newly discovered source (source 2). Note that if Source 1 and 2 form a binary system, 
the large period -- implied by the 850 AU separation -- means that proper motion will not be significant,  
and source 2 will be at the same position in the {\sc osca} and {\sc chandra} images. 
We cannot exclude the possibility that the remaining X-ray emission from Source 1, which 
is consistent with MWC~297's position, is yet 
due to one or more other companions that remain unresolved in 
the {\sc osca }image. 
Also visible in the {\sc chandra} image 
are sources 3 and 4, at roughly 35\arcsec\ from MWC~297.
Source counts were extracted using  a radius of 3~pixels for sources 1
and  2, and  a radius  of 5  pixels for  sources 3  and 4.   The total
0.3--10~keV source counts are listed in
Table~\ref{t_chan}. Note that the sum  of source counts from sources 3
and 4 is about three times greater than that for sources 1 and 2.
These {\sc chandra} data raise the question of whether the
low spatial resolution of {\sc asca} may have led to 
a misidentification of the  reported X-ray flaring of MWC~297.  We
therefore turn to the {\sc asca} flaring data.

\begin{table*}
\caption{The four sources seen in the {\sc chandra} field.}
\label{t_chan}
\begin{tabular}{cccccc}
\hline \#  & Name & RA& DEC  & counts  & $H$\\
               &      &   &      &   & \\
\hline

1 & MWC 297 & 18 27 39.54 & $-$03 49 52.0 & 23 & 4.4$^1$ \\

2 & {\sc osca} source & 18 27 39.36 & $-$03 49 49.6 & 13 & 12.9$^2$\\

3 &  2MASS J18273723-0349466 & 18 27 37.24 & $-$03 49 46.7 & 58 & 11.3$^1$ \\

4 &  2MASS J18273709-0349385 & 18 27 37.09 & $-$03 49 38.6 & 61 & 9.4$^1$ \\

\hline
\end{tabular}
\\
\noindent
$^1$ 2MASS 
$^2$ WHT/OSCA (this paper)
\end{table*}


Fig.~\ref{f_xray}~(right) shows the combined SIS
image of the {\sc asca} flaring data on MWC~297 -- on the same scale 
as the {\sc chandra} image. We have indicated the positions of 
{\sc chandra} sources 1, 3, and  4. The circles in this image have radii 
of 13\arcsec, indicating the combined {\sc asca} and {\sc chandra} 
position uncertainty. It is immediately apparent from Fig.~\ref{f_xray}~(right) 
that the {\sc chandra} sources are confused in the
{\sc asca} image. 
Moreover, the improved astrometry provided by the
corrections of Gotthelf et~al. (2000) shows that the peak of the
observed emission in the {\sc asca} PSF is inconsistent with 
the position of sources 1 and 2. Instead, the peak is consistent with 
the positions of source 3 or 4. This strongly suggests that the  
origin of the flaring behaviour is due to source 3 or 4. 
Their positions line up with 2MASS point 
sources, with $H$ = 11.3 and 9.4 for sources 3 and 4 respectively. Given the fact that Source 4 has a larger $(H-K)$ 
IR excess than Source 3, the X-ray flaring is most likely due to source 4. In any case, at
the distance of MWC~297, these $H$-band magnitudes are consistent with a 
T Tauri nature.


\section{Conclusions}
\label{s_con}

We have presented high resolution AO-NIR and X-ray imaging on MWC~297. 
Given the early spectral type of the object (B1.5), 
the reported X-ray flaring of this early Herbig Be star has been difficult to understand. 
Using {\sc chandra}, we have resolved the X-ray emission from objects surrounding MWC~297 and 
found that the brightest X-ray source is not associated with MWC~297 itself. Furthermore, we have shown that   
the peak of the observed {\sc asca} flaring is inconsistent with the position of MWC~297. 
Instead, it is most likely due to a late-type source in the Herbig Be star's field. 

The study by Stelzer et al. (2003) on late-type B stars, as well as our coronographic study 
of the young massive object MWC~297 presented here 
emphasize the need for high spatial resolution X-ray and infrared imaging 
to verify the origin of X-ray emission attributed to   
Herbig Ae/Be stars. This will assist in a proper evaluation of the physical differences governing 
low and high-mass star formation.

\begin{acknowledgements}
We thank the WHT/NAOMI team for carrying out the IR service observations, and the referee for his helpful comments.
This research has made use of the processing scripts from the Tartarus
(Version 3.0) database. JSV and PMO acknowledge financial support from PPARC.
SGE was partly supported under Marie-Curie Fellowship HDPMD-CT-2000-5
and by the Betty and Gordon Moore Foundation.
This publication makes use of data products from the Two Micron All Sky Survey (2MASS).
\end{acknowledgements}

\end{document}